\documentclass[aps,prl,reprint,noeprint,superscriptaddress]{revtex4-2}

\usepackage{graphicx}
\usepackage[colorlinks]{hyperref}
\usepackage{natbib}
\usepackage{bm}
\usepackage{mathtools}
\usepackage{color}
\usepackage{bbold}
\usepackage{lipsum}
\usepackage[utf8]{inputenc}

\newcommand{\id}{\mathbb{1}} 

\begin{document}

\title{Inverse design of mirror-symmetric disordered systems \\ 
for broadband perfect transmission}

\author{Zhazira Zhumabay}
\affiliation{Université de Rennes, CNRS, IETR - UMR 6164, F-35000 Rennes, France}

\author{Clément Ferise}
\affiliation{Université de Rennes, CNRS, IETR - UMR 6164, F-35000 Rennes, France}
\affiliation{Ecole Polytechnique Fédérale de Lausanne, Laboratory of Wave Engineering, 1015 Lausanne, Switzerland}

\author{Vincent Pagneux}
\affiliation{Laboratoire d’Acoustique de l’Université du Mans, CNRS~ UMR ~6613, Le ~Mans, France}

\author{Stefan Rotter}
\affiliation{Institute for Theoretical Physics, Vienna University of Technology (TU Wien),  A-1040 Vienna, Austria}

\author{Matthieu Davy}
\email[]{matthieu.davy@univ-rennes.fr}
\affiliation{Université de Rennes, CNRS, IETR - UMR 6164, F-35000 Rennes, France}

\date{\today}

\begin{abstract}
We present a framework for achieving broadband perfect wave transmission in complex systems by optimizing symmetric disordered media via inverse design. We show that leveraging symmetry of complex media reduces the optimization's complexity enabling the incorporation of additional constraints in the parameter space. Starting from a single perfectly transmitting state with predefined input and output wavefronts at a specific frequency, we progressively broaden the bandwidth — from a reflectionless exceptional point with a flattened lineshape to narrowband filters and ultimately to broadband quasi-perfect transmission exhibiting a rainbow effect. Numerical simulations based on the coupled dipole approximation are validated experimentally in a multichannel microwave waveguide with dielectric and metallic scatterers. Finally, we demonstrate broadband enhanced wave transmission through barriers highlighting the potential for advanced wave control applications.

\end{abstract}

\maketitle

\subsection{Introduction} 
Disordered systems are characterized by their numerous degrees of freedom and offer rich opportunities to tailor light-matter interactions \cite{Mosk2012,rotter2017light,caoShapingPropagationLight2022a}. These systems enable remarkable phenomena such as focusing of light beyond the diffraction limit \cite{Mosk2012} and achieving either perfect transmission \cite{Gerardin2014,Sarma2016} or perfect absorption of incident waves \cite{chongHiddenBlackCoherent2011,Pichler2019}. 
While conventional wavefront shaping techniques, which rely on spatial and temporal modulation of the incident wavefront, are typically restricted to a limited set of complex wavefronts within narrow frequency bands, advanced inverse design approaches of disordered matter significantly expand the scope of achievable integrated functionalities. By optimizing the spatial properties of disordered structures, these techniques allow precise control over multiple input and output wave channels in parallel. Unlike thin metasurfaces, disordered metastructures—such as multilayered metamaterials — indeed leverage multiple scattering of waves to surpass the capabilities of ordered systems \cite{Piggott2015,kwonNonlocalMetasurfacesOptical2018,sapraInverseDesignDemonstration2019,pandeSymphoticMultiplexingMedium2020a,rothammer2021tailored,yuEngineeredDisorderPhotonics2021a,liEmpoweringMetasurfacesInverse2022,caoHarnessingDisorderPhotonic2022,horodynskiAntireflectionStructurePerfect2022}. This flexibility opens new possibilities for compact, low-cost devices in microwave and photonic applications. %For disordered systems, it also facilitates the development of antireflection structures \cite{horodynskiAntireflectionStructurePerfect2022}.

%In diffusive systems, for instance, it is possible to selectively excite open and closed channels, achieving transmission near unity or zero for specific samples \cite{Gerardin2014,Sarma2016}. Additionally, lossy systems can exhibit coherent perfect absorption under the critical coupling condition \cite{Chong2010,Pichler2019,wangCoherentPerfectAbsorption2021}. However, these techniques .

Incorporating spatial symmetries into the randomness of inverse-designed systems is particularly promising for controlling transmission through obstacles. Mirror-symmetric systems inherently facilitate constructive interference between scattering paths, thereby enhancing the wave transport \cite{PhysRevB.73.165308,Whitney2009,Cheron2019,sainiMirrorSymmetryThreedimensional2024,borceaEnhancedWaveTransmission2024}. Broadband conductance enhancements have been observed in symmetric chaotic cavities or quantum dots where a central barrier is introduced \cite{Whitney2009,davyExperimentalEvidenceEnhanced2021}. In open systems, such as waveguides, symmetric disorders arranged around a central barrier can restore the bimodal distribution of transmission eigenvalues despite the presence of the obstacle leading to significantly enhanced transport across a broad frequency range \cite{Cheron2019,davyExperimentalEvidenceEnhanced2021}.

This enhancement aligns with the natural presence of reflectionless states at real frequencies in mirror-symmetric systems. Channels with zero reflection—and therefore perfect transmission in the absence of losses—occur at spectral singularities corresponding to the zeros of the reflection matrix $r(\nu)$. These zeroes are complex eigenfrequencies of a reflectionless (RL) non-Hermitian operator derived from the wave equation where incoming channels are modeled with gain and outgoing channels with loss \cite{Dhia2018,Sweeney2020,ferise2022exceptional,solReflectionlessProgrammableSignal2023,jiangCoherentControlChaotic2024}. In systems lacking symmetry, the zeros of $r(\nu)$ are generally complex making reflectionless coupling achievable only through precise optimization of the disorder. This can be accomplished either by employing inverse design techniques \cite{Piggott2015,Liu2020,horodynskiAntireflectionStructurePerfect2022} or by integrating reconfigurable elements within the system \cite{f2020perfect,delHougne2020CPA,frazierWavefrontShapingTunable2020,chen2020perfect,del2021coherent,solReflectionlessProgrammableSignal2023}. 
However, in systems with left-right mirror symmetry, the RL operator possesses a parity-time ($\mathcal{PT}$) symmetry property \cite{Dhia2018,Sweeney2020}, as it remains invariant under both a mirror transformation and the exchange of input and output channels. This symmetry constrains the zeros of  $r(\nu)$  to be either real or form complex-conjugate pairs enhancing the likelihood of real eigenfrequencies which support reflectionless coupling of incoming channels \cite{ferise2022exceptional,jiangCoherentControlChaotic2024}. When two RL eigenvalues coincide at the same frequency and the eigenvectors are coalescing, the two states form an RL-exceptional point (EP) \cite{Sweeney2019,ferise2022exceptional,wangCoherentPerfectAbsorption2021}. This phenomenon results in flattened transmission and reflection spectra, a hallmark of EP behavior \cite{Sweeney2019,wangCoherentPerfectAbsorption2021,Suwunnarat2021,sol2022meta,ferise2022exceptional,soleymaniChiralDegeneratePerfect2022,hornerCoherentPerfectAbsorption2024a}. However, the frequency range is still inherently limited, which poses challenges for practical applications requiring perfect or near-perfect transmission across broad frequency bands.

In this article, we combine the benefits of mirror-symmetry in disordered systems with inverse design techniques to create systems which achieve perfect transmission across broad frequency ranges. Our theoretical analysis reveals that the symmetry reduces the number of parameters required to place a  reflection zero on the real frequency axis, thereby simplifying the optimization process within the parameter space. This reduction in complexity also facilitates the combination of multiple zeros associated with predefined input and output wavefronts across the frequency range. Our optimization strategy involves designing a disordered medium composed of metallic and dielectric cylinders within a two-dimensional multichannel electromagnetic waveguide. We employ the coupled dipole approximation to efficiently solve the wave equation allowing rapid optimization of the structure within a limited computational time frame. We successively design i) a reflectionless exceptional point with flattened lineshape of perfect transmission; ii) bandpass filters selecting input and output wavefronts; iii) systems with flattened transmission spectra over wider frequency bands. Additionally, we extend these results to the optimization of transmission through a barrier giving transmission close to unity for predefined input wavefronts across broad frequency ranges and outperforming the case of random unoptimized symmetric disorders. Numerical simulations are supported by experimental validation in the microwave regime. 

\begin{figure*}
    \centering
    \includegraphics[width=18cm]{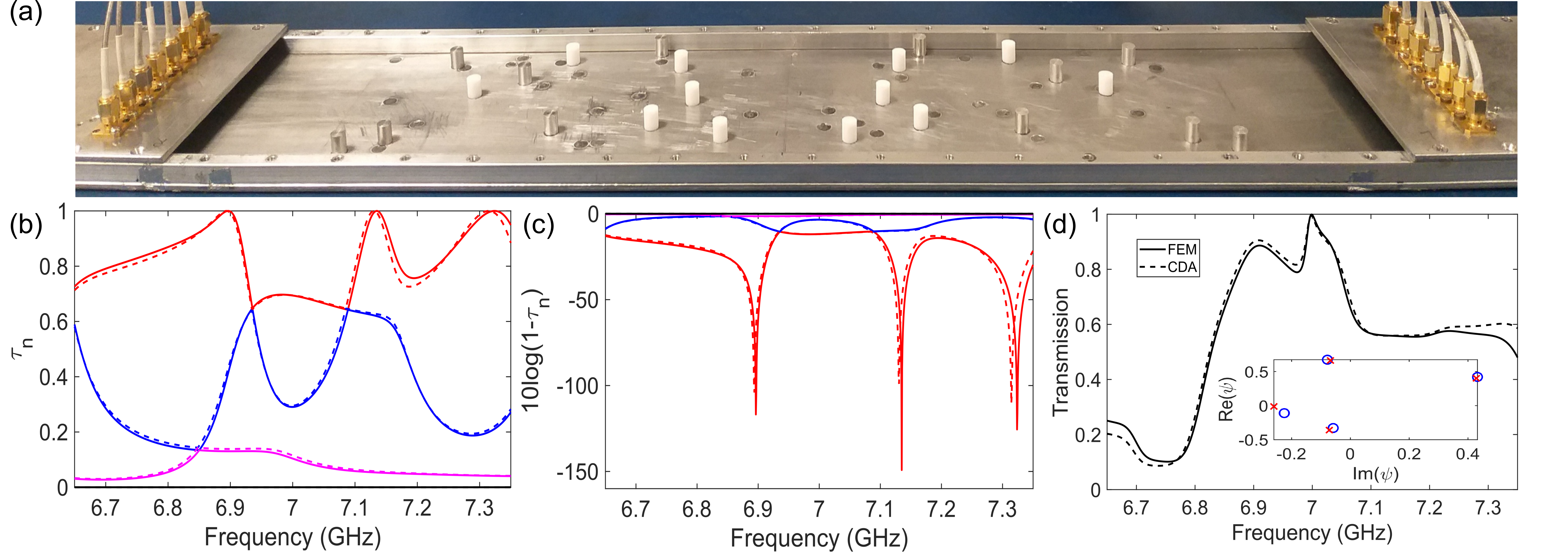}
    \caption{\textbf{Experimental setup and results of two numerical models:} (a) Photo of the experimental setup. Two arrays of seven antennas are located within a multichannel waveguide. The disorder consists of teflon and aluminum cylinders that are placed symmetrically within the waveguide. The transmission matrix in the waveguide mode space is reconstructed from measurements in the point space using a two-fold sine transformation. (b) Spectra of transmission eigenvalues $\tau_n$ of the matrix $tt^{\dagger}$ for a random symmetric disorder. Numerical simulations using a commercial software (dashed lines) are compared to numerical simulations using the coupled dipole approximation (solid lines). (c) Corresponding reflection eigenvalues $1-\tau_n$ in a dB scale. (d) Transmission through a symmetric disorder optimized for perfect transmission of a predefined input wavefront $\psi_{\mathrm{in}}$ at $\nu_0 = 7$ GHz with a predefined output wavefront $\psi_{\mathrm{out}}$. In the inset, the elements of the output wavefront $t(\nu)\psi_{\mathrm{in}}$ (red crosses) in the complex plane are compared to the elements of $\psi_{\mathrm{out}}$ (blue circles).}
    \label{fig:setup}
\end{figure*}

\subsection{Theory} 
We consider a multichannel system, such as a multichannel waveguide presented in Fig.~\ref{fig:setup}(a). The scattering matrix $S$ giving the field transmission coefficients between $2N$ coupled channels is decomposed into reflection matrices $r$ (left channels) and $r'$ (right channels), as well as corresponding transmission matrices $t$ and $t^T$, each of dimension $N \times N$. Our goal is to optimize a disorder within the medium to position at least one reflection zero on the real axis at a predetermined frequency $\nu_0$ such that $\mathrm{det}(r(\nu_0)) = 0$. In the absence of apparent symmetries, the optimization requires the optimization of the $N(N+1)/2$ complex reflection coefficients given that $r$ is symmetric. However, we demonstrate that the optimization problem is simplified in the case of a mirror-symmetric medium. For two samples with scattering matrices $S_1$ and $S_2$ placed in front of each other, the scattering matrix of the complete system $S$ can be expressed in a composite form in terms of reflection and transmission matrices of the two media, $S = S_1 \circ S_2$, where the reflection matrix is represented by \cite{horodynskiAntireflectionStructurePerfect2022}

\begin{equation}
    r = r_1 + t_1 r_2 (\id -r'_1 r_2)^{-1} t_1^T.
\end{equation}

\noindent Using the \textit{push-through identity} \cite{Sweeney2020} and the unitarity of the scattering matrix, we show in SM that for non-singular matrices $t_1$, this equation can be factorized as

\begin{equation}
    \label{eq:factorization_r}
    r = t_1^{-1} (\id - r_2 r'_1)^{-1} (r_2 - {r'}_1^{*})   t_1^{* -1}.
\end{equation}

\noindent The condition $\mathrm{det}(r(\nu_0)) = 0$ giving a real zero at frequency $\nu_0$ therefore translates into  

\begin{equation}
    \label{eq:r_zero_random}
    \mathrm{det}(r_2(\nu_0) - {r'}_1^{*}(\nu_0)) = 0.
\end{equation}

This can be interpreted as a generalized eigenvalue problem for which a vector $\psi$ must satisfy $r_2 \psi = r_1^* \psi $. %The input wavefront $\psi_{\mathrm{in}}$ giving zero reflection of the complete symmetric medium and therefore satisfying $ r(\nu_0)\psi_{\mathrm{in}} = 0$, is related to  $\psi$ by the relation $\psi_{\mathrm{in}} = t_1^{*} \psi$. Thus, $\psi_{\mathrm{in}}$ can be interpreted as the field transmitted through a virtual system of transmission matrix $t_1^{*}$ . 
This relation is reminiscent of the impedance matching \cite{imUniversalImpedanceMatching2018a} or critical coupling condition \cite{Yariv2000,Cai2000} for a single input wavefront. 

Now we focus on the case of a mirror-symmetric system. The left-right symmetry imposes a relation between reflection matrices of the two media, $r'_1 = r_2$, such that Eq.~(\ref{eq:r_zero_random}) simplifies to~\cite{schomerusScatteringTheoryComplex2013}

\begin{equation} 
    \mathrm{det}(\mathrm{Im}({r'_1(\nu_0)})) = 0.
        \label{eq:im_r}
\end{equation}
 
\noindent A reflection zero is therefore obtained when the real matrix $\mathrm{Im}({r'_1(\nu_0)})$ exhibits a zero singular value. Instead of optimizing both the real and imaginary parts of complex numbers, only the imaginary parts of $N(N+1)/2$ elements have to be optimized, thereby reducing the high-dimensional parameter space by a factor of two. The lower complexity of the optimization problem renders the zero reflection condition more accessible using inverse design techniques.  Real random matrices also exhibit reduced level repulsion between singular values compared to complex random matrices \cite{edelmanEigenvaluesConditionNumbers1988,rudelsonSmallestSingularValue2009} allowing singular values to approach zero more readily. The probability distribution of the smallest singular value $s_{\mathrm{min}}$ scales as $P(s_{\min}) \sim \exp\left(-c N s_{\min}^2\right)$, for real matrices \cite{edelmanEigenvaluesConditionNumbers1988,deanExtremeValueStatistics2008}, whereas  $P(s_{\min}) \sim s_{\min} \exp\left(-c N^2 s_{\min}^2\right)$ for complex matrices \cite{rudelsonSmallestSingularValue2009}. This makes the probability of $s_{\min} \rightarrow 0$ significantly higher for real symmetric matrices.

This reduction of the complexity in parameter space can be understood in the context of the $\mathcal{PT}$-symmetry of the non-Hermitian reflectionless operator $H_{RL}$ in systems with left-right symmetry. For symmetric perturbations, such as mirrored scatterer translations, real RL eigenvalues shift along the real axis. By optimizing the matrix $r'_1$, a reflection zero can  be tuned to a specific frequency with a minimal effort. It also enhances the ability to combine multiple zeros. Suppressing reflection for multiple incident wavefronts requires positioning multiple reflection zeros at the same real frequency. The case of zero reflection for \textit{all} incident wavefront $r = 0$ generalizing the critical coupling of a single incident wavefront to the $N$ input channels gives $r'_1 = r_2^*$~\cite{horodynskiAntireflectionStructurePerfect2022}. In the case of mirror symmetry, only the imaginary part of a reflection matrix has to be suppressed, $\mathrm{Im}({r'_1(\nu_0)})$. In addition, the linewidth of the reflection spectrum can be broadened by designing RL-exceptional points. When two eigenvalues and eigenvectors coalesce at an EP, a single input wavefront excites two reflection zeros resulting in a broadened lineshape \cite{ferise2022exceptional}.

\section{Experimental setup and numerical modeling}
\subsection{Experimental setup} 
We now demonstrate this concept using a multimode waveguide of width $W=100$~mm, length $L=400$~mm and height $h=8$~mm as shown in Fig.~\ref{fig:setup}(a) \cite{horodynskiAntireflectionStructurePerfect2022}. In the frequency range 6.5-7.5 GHz, the empty waveguide supports $N=4$ transverse modes. Since only a single mode propagates within the vertical dimension, the system can be accurately modeled using two-dimensional simulations for a scalar wavefield. The disorder consists of an ensemble of teflon cylinders with radius equal to $3.15$~mm and aluminum cylinders with radius equal to $3.075$~mm. Experimentally, we measure the transmission coefficients $\Tilde{t}$ between two arrays of 7 wire antennas positioned at a distance of 0.6~m in the waveguide with a spacing between each antenna equal to $W/8$. The 49 transmission coefficients are obtained using two ports of a Vector Network Analyzer. Each port is connected to an electro-mechanical switch which enables to select each wire antenna. The $N\times N$ transmission matrix in the mode space $t$ is subsequently reconstructed from $\Tilde{t}$ via a projection of $\Tilde{t}$ onto the waveguide modes \cite{horodynskiAntireflectionStructurePerfect2022}:

\begin{align}
\begin{split}
    t_{mn} = \sum_{y_1,y_2} &t'(y_2,y_1) \sqrt{k_n k_m}\,\text{sin}\left(\frac{m\pi}{W}y_2\right) \text{sin}\left(\frac{n\pi}{W}y_1\right),
\end{split}
\end{align}

\noindent where $y_1$ and $y_2$ are the locations of transmitting and receiving antennas, and the transverse mode number $n$ is given by $k_n = (2\pi / c_0) \sqrt{\nu^2 - (n \nu_c)^2}$ \cite{horodynskiAntireflectionStructurePerfect2022}. Because the penetration depth of the wire antennas into the waveguide is small (3~mm), they are weakly coupled to the waveguide so that a direct comparison between measurements and numerical simulations requires a renormalization of the transmission for each waveguide mode by its value for an empty waveguide. Transmission coefficients are then given by
\begin{equation}
    \Tilde{t}_{mn} =\frac{t_{mn}}{\sqrt{T^0_{n}}},
\end{equation}
where $T^0_{n}$ is the transmission through the nth mode for the empty waveguide, $T^0_{n} = \Sigma_{m=1}^N |t^0_{mn}|^2$ \cite{horodynskiAntireflectionStructurePerfect2022}.

\subsection{Optimization procedure using coupled dipole approximation}
Solving the wave equation for large structures and optimizing the positions of scatterers within them is computationally demanding. To address this challenge, we developed a model based on the coupled dipole approximation (CDA) \cite{markel2019extinction,chaumet2022discrete}, which facilitates faster computation of the scattering matrix including the case of multichannel waveguides \cite{rescanieres2024open}.

In CDA, the $N_s$ scatterers at positions $\mathbf{r}_{k}$ are characterized by an electric polarizability $\alpha_k$ which induces a dipole moment $p_k$ \cite{markel2019extinction}. The Green's function in space between a source at $\mathbf{r}_{\mathrm{in}}$ and a receiver at $\mathbf{r}_{\mathrm{out}}$, $G(\mathbf{r}_{\mathrm{out}},\mathbf{r}_{\mathrm{in}})$, is then represented as the sum of contributions from each dipole,

\begin{equation}
    \label{eq:Green_function}
    G(\mathbf{r}_{\mathrm{out}},\mathbf{r}_{\mathrm{in}}) = G_0(\mathbf{r}_{\mathrm{out}},\mathbf{r}_{\mathrm{in}}) + \Sigma_{k=1}^{N_s} G_0(\mathbf{r}_{\mathrm{out}},\mathbf{r}_{\mathrm{k}}) p_k.
\end{equation}

\noindent Rather than expressing the Green's function in spatial coordinates, our focus is on calculating the $2N \times 2N$ scattering matrix directly in the basis of flux-normalized waveguide modes. For an input mode $m$ of the waveguide and an output mode $n$, the element of the scattering matrix $S_{nm}$ is expressed in a compact form in terms of the vector of dipole moments $\mathbf{p}_m$ induced by the mth incoming waveguide mode as

\begin{equation}
    \label{eq:input_output_modes1}
    S_{nm} = S_{0nm}+ \mathbf{G}_{0n}^T \mathbf{p}_m.
\end{equation}

\noindent Here $S_0$ is the scattering matrix of the empty waveguide and $\mathbf{G}_{0n}$ is the vector of field components in an empty waveguide at dipole locations for the nth incoming waveguide mode. For an incoming mode at the left interface, $G_{0n}(\mathbf{r}_{\mathrm{k}})$ is given by 

\begin{equation}
    G_{0n}(\mathbf{r}_{\mathrm{k}}) = \sqrt{\frac{2}{W}} \mathrm{sin}(\frac{n\pi y_{\mathrm{k}}}{W})e^{i k_n x_k}.
\end{equation}

\noindent The dipole moments $\mathbf{p}_m$ is then determined by the interaction matrix $W = \mathrm{diag}(\alpha^{-1}) - G^{dd}$  as $W \mathbf{p}_m = \mathbf{G}_{0m}$, with $\mathrm{diag}(\alpha^{-1})$  the diagonal matrix of elements $\alpha_k^{-1}$. The dipole-dipole matrix $G^{dd}$ contains the Green's functions between each dipole in spatial coordinates, $G^{dd}_{i,j} = G_0(\mathbf{r}_{j},\mathbf{r}_{i}) $. The scalar Green's function of the empty waveguide between two points $\mathbf{r}_{1} = [x_1,y_1]$ and $\mathbf{r}_{2} = [x_2,y_2]$ is determined as a sum over $N_{\mathrm{eff}}$ waveguide modes,

\begin{equation}
    \label{eq:empty_Green_function}
    G_0(\mathbf{r}_{\mathrm{2}},\mathbf{r}_{\mathrm{1}}) = -\Sigma_{n=1}^{N_{\mathrm{eff}}} \frac{2}{W k_n} \mathrm{sin}(\frac{n\pi y_{\mathrm{2}}}{W}) \mathrm{sin}(\frac{n\pi y_{\mathrm{1}}}{W}) e^{i k_n |x_{\mathrm{2}}-x_{\mathrm{1}}|}.
\end{equation}

\noindent The number of waveguide modes $N_{\mathrm{eff}}$ used to compute the Green function is much larger than the number of propagating modes $N$, taking into account evanescent coupling between dipoles. 

Finally, the vector of dipole moments $\mathbf{p}_m$ is obtained from the pseudo-inverse of $W$, $\mathbf{p}_m = W^{-1} \mathbf{G}_{0m}$, and Eq.~(\ref{eq:input_output_modes1}) yields $S_{nm} = S_{0nm}+ \mathbf{G}_{0n}^T W^{-1} \mathbf{G}_{0m}$. The scattering matrix in the basis of flux-normalized waveguide modes is therefore expressed as

\begin{equation}
    \label{eq:input_output_modes}
    S = S_{0}+ {G}_{0}^T W^{-1} {G}_{0},
\end{equation}

\noindent with ${G}_{0}$ being the $N_s \times 2N$ matrix of vectors $\mathbf{G}_{0n}$. 

The expression of $W$ includes the return Green's functions $G_0(\mathbf{r}_{k},\mathbf{r}_{k})$ at scatterer locations, which are the diagonal elements of $G^{dd}$. They ensure the unitarity of $S$ \cite{markel2019extinction}. The same results can be obtained by renormalizing the bare polarizability ${\alpha}_k$ to $\Tilde{\alpha}_k = \alpha_k/(1-G_0(\mathbf{r}_{\mathrm{k}},\mathbf{r}_{\mathrm{k}})\alpha_k)$ \cite{markel2019extinction} and setting the diagonal elements of $G^{dd}$ to zero. We note that the bare polarizabilities for our lossless scatterers are imaginary numbers that are positive for dielectric scatterers and negative for metallic scatterers. Moreover, the return Green's function displays a logarithmic divergence as $N_{\mathrm{eff}} \rightarrow \infty$ since $G_0(\mathbf{r}_{i},\mathbf{r}_{i}) \propto \Sigma_{n=1}^{N_{\mathrm{eff}}} k_n^{-1}$ \cite{riveroGreensFunctionDefect2021}. We therefore apply a truncation on the sum, $N_{\mathrm{eff}} = 6N$. Finally, as the radar cross-section of metallic cylinders is large, they cannot be modeled with a single dipole. We therefore mimick a metallic scatterer by eight dipoles placed on its boundary. The values of $\alpha$ for metallic and dielectric cylinders are extracted from a calibration test by comparing $S$ obtained from the model to $S$ obtained from a full-wave commercial simulation software (COMSOL Multiphysics), see SM. 

\begin{figure*}
    \centering
    \includegraphics[width=18cm]{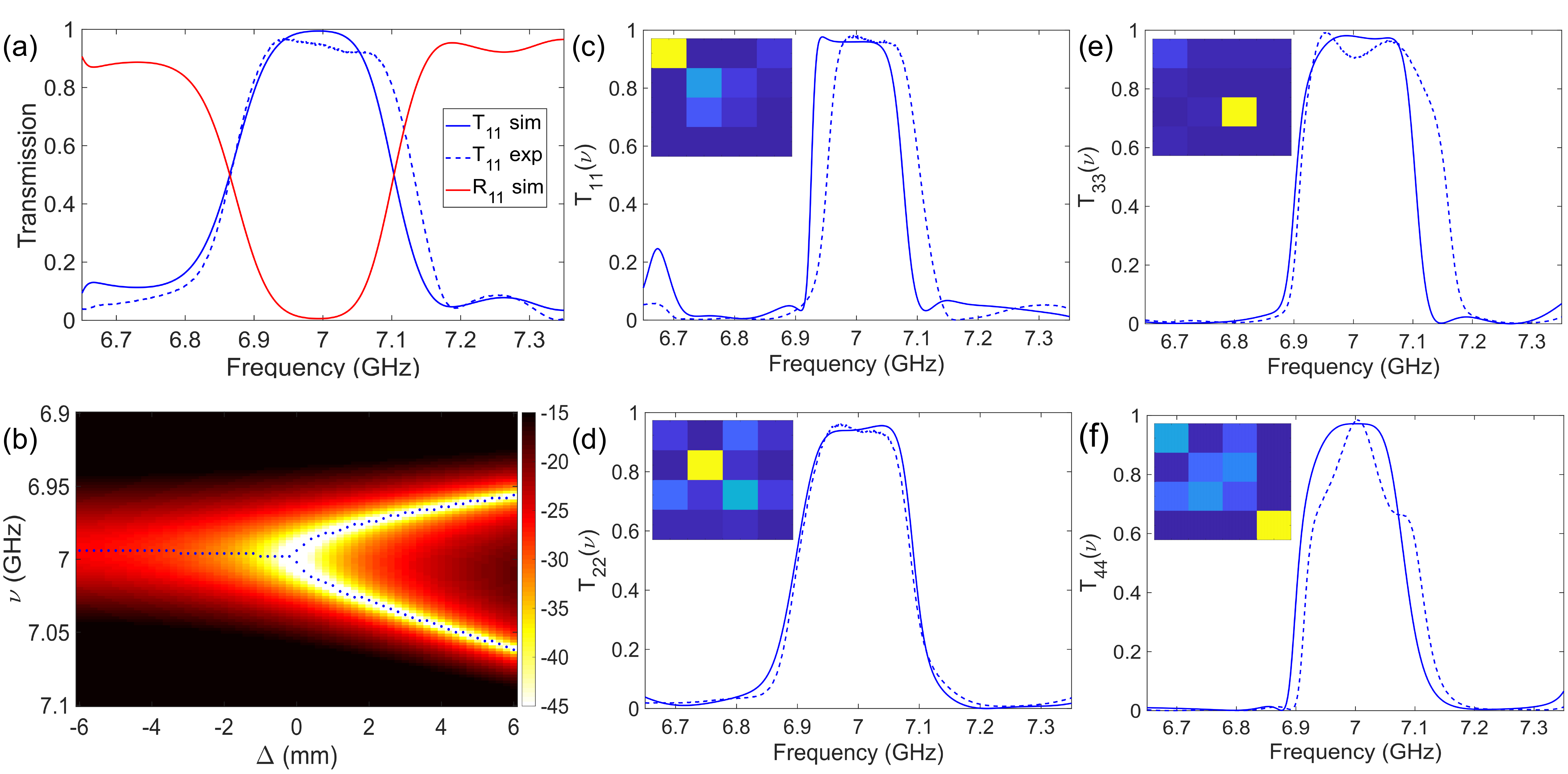}
    \caption{\textbf{Reflectionless Exceptional Point and bandpass filtering:} (a) Spectra of the amplitudes $T_{11}$ and $R_{11}$ obtained from simulations (blue and red solid lines) and experimentally (blue dashed line). (b) The reflection for the first waveguide mode in a dB scale $-10 \mathrm{log10}(R_{11})$ obtained numerically is shown as a function of frequency and a symmetric displacement $\Delta$ of a teflon scatterer on each side of the symmetric medium. The presence of a RL-EP is evidenced by the presence of a single peak for $\Delta<0$ and two peaks for $\Delta>0$ (dashed blue line). (c-f) Spectra of transmission $T_{nn}(\nu)$  for bandpass filters selecting a single input mode and a single output mode with a transmission matrix measured at 7 GHz in the insets.}
    \label{fig:EP}
\end{figure*}

To verify the accuracy of the CDA model, we start with a random disorder with left-right symmetry. Specifically, 5 teflon cylinders and 6 aluminum cylinders are randomly placed on each side, resulting in a total of 10 teflon cylinders and 12 aluminum cylinders. The unitarity error, defined as $\epsilon = \| SS^\dagger - \id \|_2^2$ remains below $10^{-9}$. Spectra of transmission eigenvalues $\tau_n$ found from a diagonalization of the matrix  $tt^\dagger$ are shown in Fig.~\ref{fig:setup}(b). Reflection eigenvalues $1-\tau_n$ are presented in a dB scale in Fig.~\ref{fig:setup}(c). CDA simulations are in excellent agreement with simulations obtained numerically with a commercial software (COMSOL). The left-right symmetry results in the presence of three reflectionless states with the first transmission eigenvalue $\tau_1$ reaching unity. Correspondingly, the smallest reflection eigenvalue $1-\tau_n$ shown on a log-scale in Fig.~\ref{fig:setup}(c) exhibits three pronounced dips with reflection below -100 dB. 

Because the model only requires computing the inverse of a matrix of dimension equal to the number of dipoles, it is particularly adapted to large-scale problems. The full-wave computation of the $S$-matrix with the commercial numerical software typically takes $\sim 15$ s for a single frequency. Even though the gradient of the cost function can be calculated directly from the field inside, solving complex optimization problems requiring hundreds of iterations would take hours. In contrast, the calculation time of a single $S$-matrix is reduced to a few milliseconds within the framework of the CDA so the optimization problems can be solved within a few minutes. 

\section{Optimization results} 

\subsection{Perfect transmission with predefined input and output wavefronts at selected frequency} 
We first design a symmetric disorder for which perfect transmission of a single channel is obtained at a prescribed frequency $\nu_0 = 7$~GHz. We also add another constraint: input and output wavefronts of perfect transmission must correspond to predefined vectors $\psi_{\mathrm{in}}$ and  $\psi_{\mathrm{out}}$. 
Adapting the disorder to match an imposed arbitrary wavefront rather than shaping the incident wavefront as traditionally done in wavefront shaping eliminates the need for precise manipulation of the incident wavefront \cite{del2021coherent}. We first define the elements of the incident vector $\psi_{\mathrm{in}}$ to be all real and equal to 0.5. This corresponds to the same excitation through all waveguide modes. The elements of the output wavefront $\psi_{\mathrm{out}}$ are chosen to be predefined random complex numbers. Since we expect perfect transmission, both vectors are normalized to unity $||\psi_{\mathrm{in}}|| = ||\psi_{\mathrm{out}}|| = 1$. 

We optimize the positions of the cylinders using a gradient-descent algorithm based on the ADAM optimizer \cite{kingmaAdamMethodStochastic2014}. The cost function is defined as $f = \|t(\nu_0)\psi_{\mathrm{in}} - \psi_{\mathrm{out}} \|^2$, where $t(\nu_0)$ is the transmission matrix of the mirror-symmetric medium at frequency $\nu_0$. We initiate the optimization with a randomly generated symmetric disorder. Throughout the optimization process, the scatterers on the right side are shifted to align with the positions of those on the left, ensuring the system maintains its mirror symmetry. In practice, it suffices to compute the scattering matrix $S_1(\nu)$ of the left medium to determine the complete scattering matrix $S = S_1 \circ S_2$. The scattering matrix of the right medium $S_2$ is indeed easily found from $S_1$ using the symmetry relations $r_2 = r'_1$, $t_2 = t_1^T$ and $r'_2 = r_1$. The transmission matrix of the complete medium is therefore now $t = t_1^T (\id - {r'_1}^{2}) t_1$ \cite{horodynskiAntireflectionStructurePerfect2022}. While Eq.~(\ref{eq:im_r}) shows that $\mathrm{Im}(r'_1)$ needs to be adjusted to get perfect transmission of the incident wavefront $\psi_{\mathrm{in}}$, fixing the output wavefront also requires adjusting the transmission matrix $t$ and thus $t_1$.

At the end of the optimization, the spectrum of the transmission $T(\nu)=||t(\nu)\phi_{\mathrm{in}}||^2$ presents a peak at $\nu_0$, $T(\nu) = 0.999$, as seen in Fig.~\ref{fig:setup}(d). The complex elements of the output wavefront at $\nu_0$, $t(\nu)\phi_{\mathrm{in}}$, closely correspond to the predefined one $\psi_{\mathrm{out}}$ (see the inset of Fig.~\ref{fig:setup}(d)). This validates our optimization strategy to control both the position of a reflectionless state together with the corresponding input and output wavefronts.

\subsection{On-demand reflectionless exceptional point}
After having established that a reflection zero can be placed at a predefined frequency with precise control on input and output wavefronts, we now extend the capabilities of our inverse design technique to  achieve broadband quasi-perfect transmission of predefined input modes. We first turn to the case of a reflectionless exceptional point (RL-EP) corresponding to two RL-states coinciding at the same real eigenfrequency. The hallmark of an RL-EP is indeed a flattening of the reflectionless coefficient around $\nu=\nu_0$, where the reflection  exhibits a quartic lineshape $|r(\nu) \phi_{\mathrm{in}}|^2 \propto (\nu-\nu_0)^4$, as opposed to the quadratic dependence observed for a single RL-state \cite{Sweeney2019, ferise2022exceptional}. 

We create an RL-EP corresponding to the first waveguide mode by optimizing the transmission over a frequency range of $\delta\nu = 100$~MHz around $\nu_0 = 7$~GHz. The cost function is hence  

% \begin{equation}
\begin{equation}
    f = \int_{\nu_0-\delta \nu/2}^{\nu_0+\delta\nu/2} [1-T_{11}(\nu)]d\nu,
\end{equation}
% \end{equation}

\noindent where $T_{11}(\nu) = |t_{11}(\nu)|^2$ is transmission between first input mode and first output mode. The integral is estimated using 5 frequency points. The spectra of  transmission and reflection $T_{11}$ and $R_{11}$ are presented in Fig.~\ref{fig:EP}(a). Their shapes are noticeably flattened near $\nu_0$ in comparison with a single RL zero shown in Fig.~\ref{fig:setup}(d). The experimental transmission measurements are in very good agreement with these numerical findings. 

To verify that the flattened shape arises from the presence of an RL-EP, we slightly translate two mirror-symmetric teflon cylinders on either side while preserving the system's symmetry. The reflection $R_{11,dB}$ in a dB scale is shown in Fig.~\ref{fig:EP}(b) and its local minima are extracted for each translation distance $\Delta$. For $\Delta >0$, the two local minima in $R$ correspond to two RL eigenvalues positioned on the real axis at different frequencies. These eigenvalues converge to an exceptional point at $\nu_0$ as $\Delta $ approaches 0 and then leave the real axis in a conjugate pair for $\Delta < 0$, leading to a single minimum in $R$ with increasing reflection \cite{ferise2022exceptional}. This behavior fully confirms that the spectrum corresponds to two RL zeroes coalescing at an EP at $\nu_0$. 

\begin{figure*}
    %\centering
    \includegraphics[width=18cm]{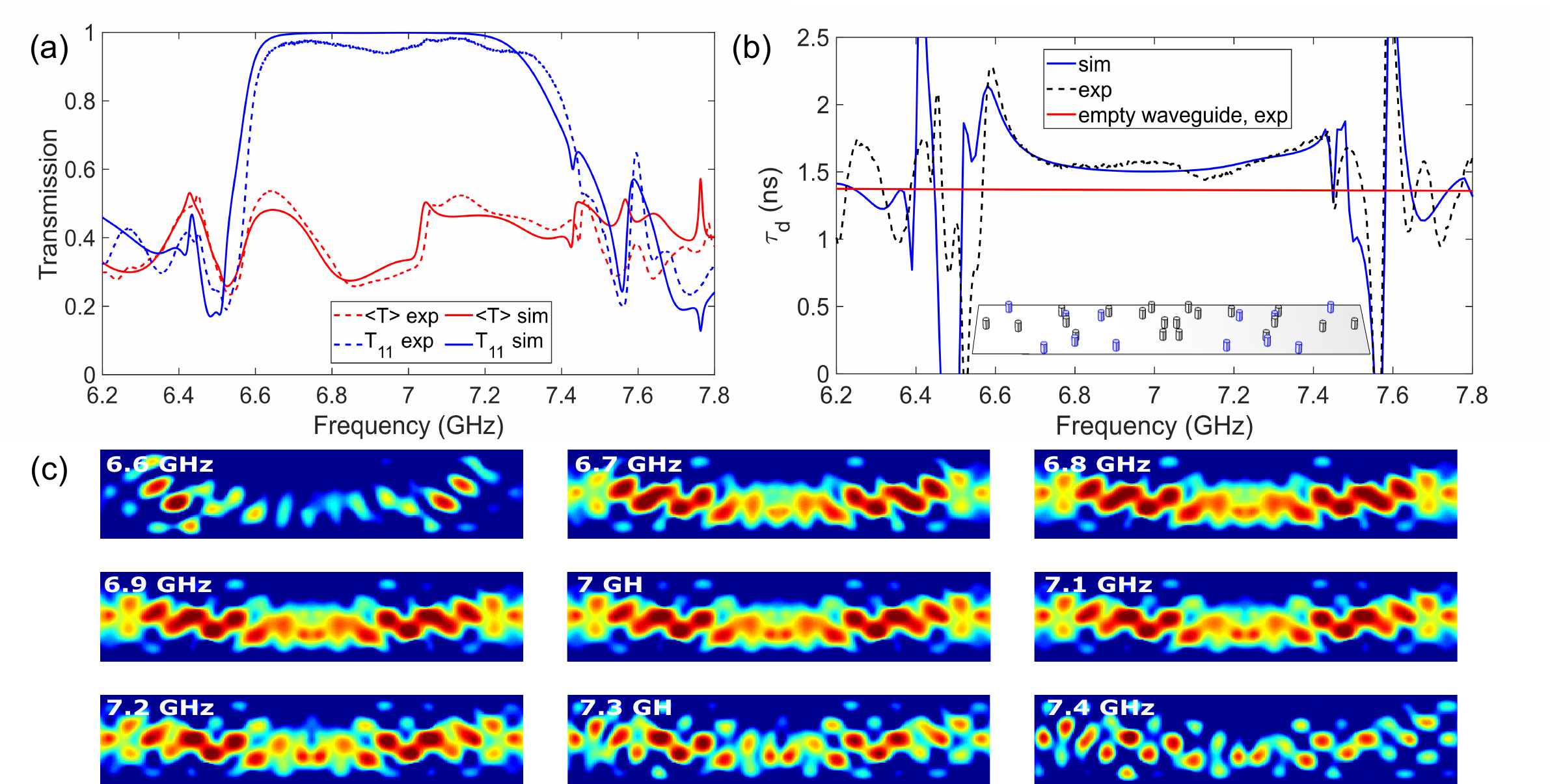}
    \caption{\textbf{Rainbow-trapping effect of transmission of the first mode}. (a) Spectrum of $T_{11}(\nu)$ (blue line) and mean transmission $T(\nu)$ (red line) from simulations (solid lines) and experiments (dashed lines). The transmission of the first mode was optimized over a bandwidth of 600~MHz around $\nu_0 = 7$~GHz. (b) Single channel phase delay time of the first mode, $\tau_{d} = d\phi_{11} / d\omega$. In the inset, a sketch of the optimized configuration with aluminum (blue) and teflon (dark gray) cylinders. (c) Maps of the energy density inside the waveguide at nine different frequencies. Between 6.7~GHz and 7.3~GHz the complex pattern is seen to be highly correlated.}
    \label{fig:rainbow}
\end{figure*}

\subsection{Bandpass single mode filtering}
To go beyond the flattened lineshape of an RL-EP, our objective is now to develop a single-mode bandpass filter that maintains input and output wavefronts minimizing distortion and ensuring high signal fidelity. Bandpass filters play a crucial role in microwave and optical signal processing, while mode multiplexing is essential for preserving transmission quality in high-precision applications \cite{wu2008balanced, ho2013linear}. These filters enable quasi-perfect transmission across a broad frequency range while effectively suppressing unwanted spectral components outside this band \cite{wu2008balanced}. In contrast, an RL-EP extends quasi-perfect transmission over a wider band but it does not address behavior beyond this range. 

We optimize the symmetric scattering medium for quasi-perfect transmission within $\delta\nu = 200$~MHz around $\nu_0 = 7$~GHz while ensuring near-zero transmission outside this range over a $\Delta \nu = 600$~MHz span. The input and output wavefronts are identical corresponding to a specific waveguide mode. For the nth mode, the cost function is defined as follows: 
\begin{multline}
    f = \int_{\nu_0-\Delta \nu/2}^{\nu_0-\delta\nu/2} T_{nn}(\nu)d\nu + \int_{\nu_0-\delta \nu/2}^{\nu_0+\delta\nu/2} [1-T_{nn}(\nu)]d\nu \\
    + \int_{\nu_0+\delta \nu/2}^{\nu_0+\Delta\nu/2} T_{nn}(\nu)d\nu.
\end{multline}

In practice, the integral is approximated by sampling 15 frequency points within the designated range. The optimization is performed separately for each input waveguide mode, yielding four distinct configurations that enable bandpass filtering for each mode.  Figure~\ref{fig:EP}(c–f) presents the numerical optimization results for these four filters alongside the corresponding experimental data.  The transmission matrices in the insets confirm that transmission is fully focused within the selected mode while being suppressed in the others. In all cases, the simulation and measurement spectra exhibit strong agreement. However, discrepancies arise near the passband edges, where the sharp transmission drop increases sensitivity to experimental conditions. Our approach not only enhances spectral selectivity but also enables robust mode management, paving the way for the design of low-loss, compact filters for advanced applications in telecommunications, sensing, and quantum information processing.

\subsection{Ultra-broadband quasi-perfect transmission} 
While the bandwidth of the results presented above is limited to approximately 100 MHz, we further expand our approach by optimizing the medium to achieve quasi-perfect transmission of the first waveguide mode across a much broader frequency range, here 600 MHz spanning from 6.7 GHz to 7.3 GHz. This range significantly exceeds the spectral correlation width typically observed in random disordered systems with a comparable number of scatterers. No restrictions are included outside this band. The optimization results and corresponding measurements are shown in Fig.~\ref{fig:rainbow}(a). The transmission $T_{11}(\nu)$ forms a plateau over the targeted bandwidth. The smallest reflection eigenvalue $1-\tau_1(\nu)$ shown in a dB scale %(see SM)
reveals that the flat shape is due to the presence of three reflection zeroes. For these reflectionless states, the input and output wavefronts are strongly correlated with the first waveguide mode leading to quasi-perfect transmission over the targeted range. 

This phenomenon is reminiscent of broadband rainbow-trapping absorption observed in acoustics for broadband ultrasound absorption \cite{jimenez2017rainbow,al2018coupled} and for water waves \cite{wilks2022rainbow}. In these cases, critical coupling of incident waves is achieved at multiple successive frequencies by loading a single-channel waveguide with Helmholtz resonators \cite{jimenez2017rainbow,al2018coupled}. The overlapping of low-quality-factor resonances produces a broadband quasi-perfect absorption spectrum \cite{jimenez2017rainbow,al2018coupled}. However, a key distinction exists between the two mechanisms. Unlike rainbow-trapping absorption, where the field at each resonance is predominantly localized within a single resonator, our system exhibits remarkable stability in the intensity distribution throughout the sample. 

The single channel phase delay time and the energy density maps indicate that the optimization leads to a complex scattering process guiding the first mode like a particle-like state through the disorder within the band \cite{rotterGeneratingParticlelikeScattering2011,gerardinParticlelikeWavePackets2016,bohmSituRealizationParticlelike2018}, while scattering outside this band reduces transmission. The single channel phase delay time, $\tau_{d} = d\phi_{11} / d\omega$, where $\phi_{11}$ is the phase of the transmission coefficient $t_{11}$, is seen to be flat within the bandpass (shown in Fig.~\ref{fig:rainbow}(b)), as it is the case of ballistic propagation through an empty waveguide. Moreover, numerical simulations of the energy density maps reveal a striking similarity across different frequencies within the targeted bandwidth (see Fig.~\ref{fig:rainbow}(c)). The presence of multiple reflectionless states with correlated input and output wavefronts translates into a strong correlation of the field distribution within the sample as well.

\begin{figure*}
    \centering
    \includegraphics[width=18cm]{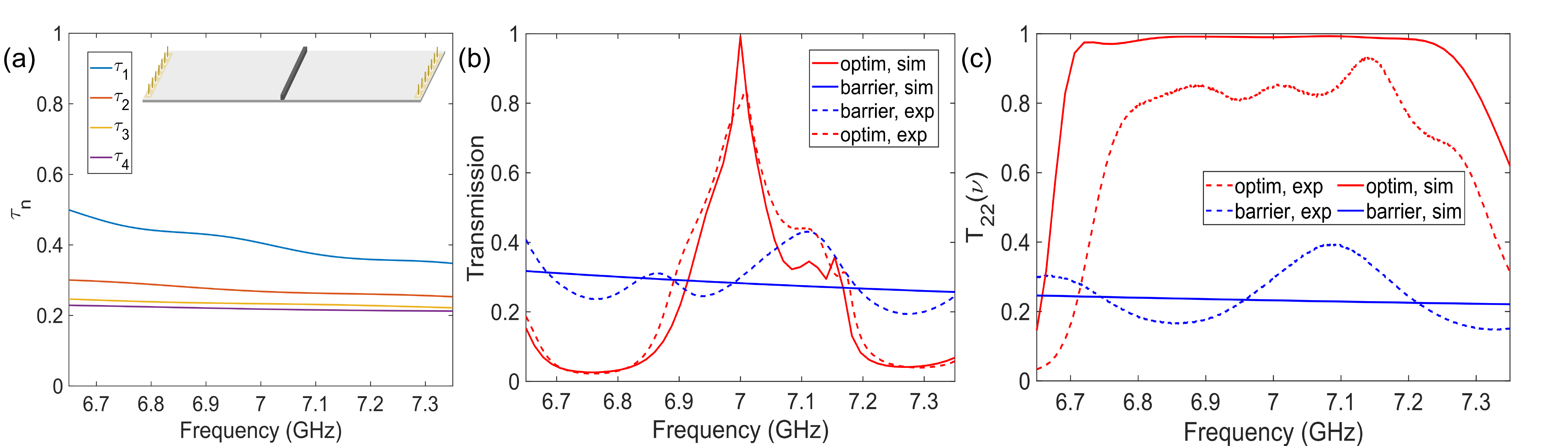}
        \caption{\textbf{Maximizing the transmission around a barrier by means of symmetric disorders}. (a) Numerical spectra of transmission eigenvalues $\tau_n$ for a barrier placed in the middle of the multichannel waveguide. (b) Numerical (solid lines) and experimental (dashed lines) spectra of transmission averaged over input channels $T(\nu)$ for a symmetric disorder optimized to get perfect transmission of all incident channels at $\nu_0 = 7$~GHz. (c) Spectra of transmission $T_{22}$ for a configuration giving maximal transmission of the second waveguide mode between 6.65~GHz and 7.35~GHz.}
    \label{fig:barrier}
\end{figure*}

\subsection{Optimization of transmission through a barrier}
We finally explore the enhancement of transmission through an obstacle, which is a barrier placed inside the medium. The barrier is a metallic bar covering the whole waveguide width $W$ and a square cross-section $h_b^2 = 6 \times 6$ $\mathrm{mm}^2$. As the height $h_b$ is smaller than the waveguide height $h = 8$ mm, the bar can be modeled by a dielectric barrier in a two-dimensional geometry. We first perform a three-dimensional numerical simulation of the $8\times 8$ scattering matrix $S_b(\nu)$ of the barrier. Even though the system is mirror-symmetric, wavefront shaping cannot counteract the reflection on the barrier. As can be seen from Fig.~\ref{fig:barrier}(a), the maximum transmission eigenvalue is equal to $\tau_1 = 0.5$, meaning that the RL-eigenvalues only come as conjugate pairs, none of them being on the real axis. However, as demonstrated both numerically and experimentally, transmission can be significantly enhanced on a broadband range by introducing a mirror-symmetric disorder around the barrier \cite{Cheron2019,davyExperimentalEvidenceEnhanced2021}. The transmission through the barrier alone averaged over waveguide modes, $T(\nu_0) = N^{-1} \Sigma_{n=1}^4 T_n(\nu_0)$ at $\nu_0$ is equal to 0.28 at $\nu_0$ (see Fig.~\ref{fig:barrier}(b)).

%The disorder shifts the RL-eigenvalues within the complex plane, causing a fraction of the complex eigenvalues to coalesce at exceptional points on the real axis and subsequently become real. These real RL-eigenvalues at specific frequencies, along with those near the real axis, give rise to open transmission eigenchannels with transmission close to unity. The resulting enhancement in transmission is broadband, as it does not depend on a single resonance but instead leverages all RL-states.

To optimize the mirror-symmetric disorder around the barrier,  $S_b(\nu)$ is used in the coupled dipole approximation. The total scattering matrix is constructed from  $S = S_L \circ S_b \circ S_R$, where the scattering matrices $S_L$ and $S_R$ correspond to the mirror-symmetric regions on left and right sides of the barrier.
We then optimize a symmetric disorder made of 5 metallic cylinders on each side of the barrier to achieve perfect transmission of \textit{all} incident channels at single frequency $\nu_0$. This anti-reflection structure requires each of the $N$ input channels to have transmission equal to unity \cite{horodynskiAntireflectionStructurePerfect2022}. This condition means that $N$ reflection zeroes lie at $\nu_0$ on the real axis and that the corresponding incident wavefronts are orthogonal. The average transmission of the optimized sample is illustrated in Fig.~\ref{fig:barrier}(b). Numerically, $T(\nu_0)$ is equal to 0.99 with an enhancement by a factor of $\sim 3$ relative to the absence of disorders around the barrier. Experimentally, transmission is slightly reduced due to losses within the waveguide, $T(\nu_0) = 0.82$. 

Next, we combine multiple zeroes for broadband transmission of a single channel through the barrier. The result for the second mode and a frequency range of 450~MHz is shown in Fig.~\ref{fig:barrier}(c). Similar results are presented in the SM for the other modes. In all cases, the transmission is flat over the target band with numerical results accurately reproduced by experimental ones. This demonstrates that the optimization of mirror-symmetric systems can provide quasi-perfect and broadband transmission not only when the entire medium is optimized, but also when an immovable obstacle is placed at the center of the medium. 

\section{Conclusion}
In this work, we have demonstrated an inverse design approach for achieving broadband perfect transmission in mirror-symmetric disordered systems. We significantly reduced the complexity of the optimization process enabling efficient control over multiple reflectionless states by leveraging the symmetry constraints. Through numerical simulations and experimental validation in a multichannel microwave waveguide, we have successfully designed reflectionless exceptional points, bandpass filters, and broadband quasi-perfect transmission states. These results highlight the ability of mirror-symmetric disorders to enhance wave transport beyond the capabilities of conventional random media.
A key finding is the ability to multiplex multiple reflection zeros at real frequencies facilitating quasi-perfect transmission over extended frequency ranges. Unlike traditional wavefront shaping techniques, which require precise control of the incident wavefront, our approach optimizes the disorder itself ensuring robust transmission without active modulation. 

Finally, we demonstrated that mirror-symmetric scattering structures can be used to enhance transmission through barriers presenting a promising strategy for overcoming wave-blocking obstacles in complex environments. In the future, we envisage that different mechanisms could be used to reconfigure the environment in situ, while maintaining symmetry. This could involve varying the height of the cylinders using miniaturized linear actuators in a two-dimensional geometry \cite{faul2024agile} or introducing reconfigurable intelligent surfaces into three-dimensional environments \cite{del2021coherent}. 

These results pave the way for advanced wave control applications, including high-fidelity signal transmission, compact multiplexing devices and efficient energy transport in photonic and microwave systems. Future research could explore the extension of these concepts to three-dimensional systems, nonlinear media \cite{wangNonlinearityinducedScatteringZero2024,goicoecheaDetectingFocusingNonlinear2024}, and reconfigurable structures, further broadening the scope of inverse-designed disordered symmetric materials in wave physics and engineering.

\section{acknowledgements}
\begin{acknowledgements}
This work is supported in part by the European Union through European Regional Development Fund (ERDF), Ministry of Higher Education and Research, CNRS, Brittany region, Conseils Départementaux d’Ille-et-Vilaine and Côtes d’Armor, Rennes Métropole, and Lannion Trégor Communauté, through the CPER Project CyMoCod, in part by the French ``Agence Nationale de la Recherche" (ANR) under Grant ANR-24-CE91-0007-01 for the project META-INCOME and by the Austrian Science Fund (FWF) [10.55776/PIN7240924]. For open access purposes, the author has applied a CC BY public copyright license to any author-accepted manuscript version arising from this submission. M.D. acknowledges the Institut Universitaire de France. 
\end{acknowledgements}

% \bibliography{WS_NL}
%\bibliographystyle{apsrev4-1}
%\bibliography{bibliography.bib}% Produces the bibliography via BibTeX.

%Control: production of eprint (1) enabled
\providecommand{\noopsort}[1]{}\providecommand{\singleletter}[1]{#1}%
%

%%%%%%%%%% Merge with supplemental materials %%%%%%%%%%
\widetext
\begin{center}
\textbf{\large Supplemental Material for ``Inverse design of mirror-symmetric disordered systems for broadband perfect transmission"}
\end{center}
%%%%%%%%%% Merge with supplemental materials %%%%%%%%%%
%%%%%%%%%% Prefix a "S" to all equations, figures, tables and reset the counter %%%%%%%%%%
\setcounter{equation}{0}
\setcounter{figure}{0}
\setcounter{table}{0}
\setcounter{page}{1}
\makeatletter
\renewcommand{\theequation}{S\arabic{equation}}
\renewcommand{\thefigure}{S\arabic{figure}}
\setcounter{secnumdepth}{4}
%%%%%%%%%% Prefix a "S" to all equations, figures, tables and reset the counter %%%%%%%%%%

\section{Demonstration of the factorized form of the reflection matrix}

For two complex media with scattering matrices $S_1$ and $S_2$, the scattering matrix of the complete system can be written in a composite form \cite{horodynskiAntireflectionStructurePerfect2022}. In particular, the reflection matrix is expressed by

\begin{equation}
    r = r_1 + t_1 r_2 (\id -r'_1 r_2)^{-1} t_1^T.
\end{equation}

Then we use the push-pull relation $(A+BC)^{-1} B = A^{-1} B (\id + CA^{-1} B)^{-1}$ with $A=\id$, $B=t_1^T$ and $C=-t_1^{T-1} r'_1 r_2$. We also use the fact that the scattering matrix $S_1$ is unitary $S_1{^\dagger} S_1 = \id$ giving $r_1 t_1^{T-1} = -t_1^{\dagger -1}{r'}_1^{\dagger}$ and ${r'}_1^{\dagger} r'_1 + t_1^{\dagger} t_1 = \id$. This gives
\begin{equation}
    r = [r_1 -r_1 t_1^{T-1}r'_1 r_2 t_1^T+t_1 r_2 t_1^T][1-t_1^{T-1}r'_1r_2t_1^T]^{-1}.
\end{equation}

We denote now $X=r_1 -r_1 t_1^{T-1}r'_1 r_2 t_1^T+t_1 r_2 t_1^T$. Then we use the fact that $r_1 t_1^{T-1} = -t_1^{\dagger -1}{r'}_1^{\dagger}$ to get
\begin{equation}
    X = [-t_1^{\dagger -1}{r'}_1^{\dagger}+(t_1^{\dagger -1}{r'}_1^{\dagger}r'_1+t_1)r_2]t_1^{T}
\end{equation}

and

\begin{equation}
    X = t_1^{\dagger -1} [-{r'}_1^{\dagger}+({r'}_1^{\dagger}r'_1+t_1^{\dagger }t_1)r_2]t_1^{T}.
\end{equation}

The identity ${r'}_1^{\dagger}r'_1+t_1^{\dagger }t_1 = \id$ leads to $X = t_1^{\dagger -1} [-{r'}_1^{\dagger}+r_2]t_1^{T}$. This finally leads to the factorization

\begin{equation}
    r = t_1^{\dagger -1} (r_2 - {r'}_1^{\dagger}) (\id - r'_1 r_2)^{-1} t_1^{T -1}.
\end{equation}

Finally, since the matrix $r$ is symmetric, its transpose leads to Eq.~2 of the main text. Note that if $t_1$ is non-invertible, the factorization cannot be written since $t_1^{-1}$ is not defined. In this case, a closed channel with zero transmission and perfect reflection exists. This channel is therefore unaffected by the presence of the second medium.

\section{Calibration test to find the polarizability of the dipoles}

\begin{figure}
    \centering
    \includegraphics[width=0.5\textwidth]{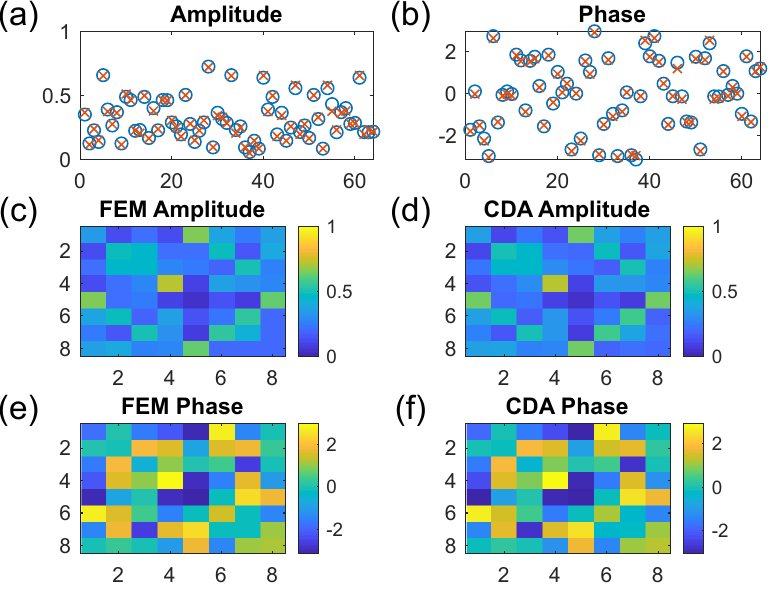}
    \caption{\textbf{Results of the calibration test to find bare polarizabilty of the cylinders.} \textbf{(a)-(b):} Amplitude (a) and phase (b) of the $N^2 = 64$ elements of the scattering matrices computed in FEM (blue circles) and CDA (red crosses). \textbf{(c)-(f):} Corresponding colormaps of the amplitude (c,d) and phase (e,f).}
    \label{fig:calibration}
\end{figure}

\begin{figure*}
    \centering
    \includegraphics[width=18cm]{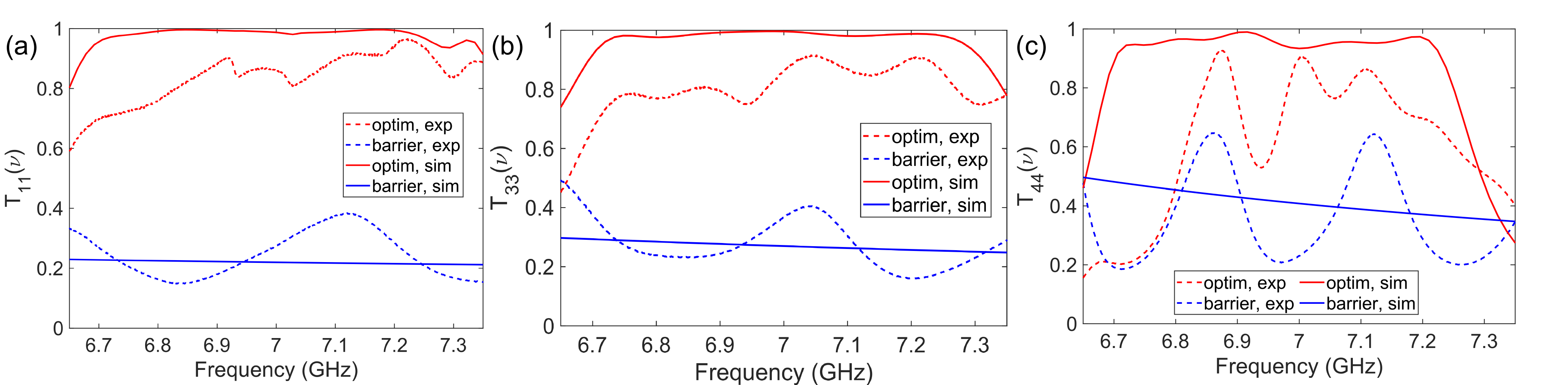}
        \caption{\textbf{Broadband optimization of transmission through a barrier for waveguide modes 1, 3 and 4 (a)-(c)}. Spectra of transmission for an optimized configuration giving maximal transmission of the first (a), third (b) and fourth (c) waveguide mode between 6.65~GHz and 7.35~GHz.}
    \label{fig:barrierSM}
\end{figure*}

We use the coupled dipole approximation (CDA) to simulate the scattering matrix of a disordered waveguide. To do so, each scatterer within the waveguide is described by a collection of dipoles with polarizability $\alpha_n$. Because metallic cylinders have a large radar cross section, they cannot be modeled by a single dipole. We instead use eight dipoles placed on its boundary, each of them having the same polarizability. To extract the proper value of $\alpha_n$ at 7 GHz, we first simulate the scattering matrix corresponding to a single scatterer randomly placed within the multichannel waveguide using the commercial software COMSOL based on the finite element method (FEM), giving $S_{fem}$. The waveguide supports $N=4$ waveguide modes at the given frequency. We find the polarizability $\alpha_n$ which minimizes the norm of the difference between $S_{fem}$ and the scattering matrix provided by the CDA code $S_{CDA}$: $f = \| S_{CDA}-S_{fem} \|^2$. The obtained polarizability is an imaginary number equal to $\alpha_n = -6i$ for an aluminum cylinder and $\alpha_n = 0.048i$ for a teflon cylinder. Both cylinders have a radius around $r =3.1$~mm. Given the definition of the Green's functions provided in the main text, $\alpha_n$ is a negative imaginary number for metallic cylinders and a positive imaginary number for dielectric ones within our frequency range. For lossless particles, the real part of $\alpha_n$ is zero. 

To verify the accuracy of our model, we then compared simulations for 10 aluminum cylinders randomly placed in the waveguide. The minimal distance between the centers of two cylinders is equal to $3r$ since a requirement of CDA is that the dipoles do not approach too closely \cite{balla2012coupled, markel2019extinction}. The amplitudes and phases of the $N^2=64$ of the scattering matrix are shown in Fig.~\ref{fig:calibration}. An excellent agreement is observed. We also find that the variations of $\alpha_n$ over the bandwidth considered in simulations are small.

% \section{Results in a logarithmic scale}

% Fig.~6 shows reflection in a dB scale of the broadband transmission and the two reflection eigenvalues of the optimized configuration from Fig.~3, which demonstrated rainbow-trapping effect of transmission inside the waveguide. The three reflection zeroes have been found in that case, as it is seen from the first reflection eigenvalue. Additionally, the second reflection eigenvalue displays several dips with reflection below -100dB. 

% \begin{figure}
%     \centering
%     \includegraphics[width=0.5\textwidth]{rb_db.pdf}
%     \caption{\textbf{Spectra of reflection $R_{11}$ and the two first reflection eigenvalues:} The presence of three reflection zeroes is found.}
% \end{figure}

\section{Maximizing the transmission around a barrier for other three modes}
In the main text, we demonstrate the quasi-perfect transmission of the second waveguide mode through a barrier by optimizing the surrounding symmetric disorder. Expanding on these findings, we now present results for three additional waveguide modes. Fig.~\ref{fig:barrierSM}(a) shows the transmission spectra for the optimized configuration giving maximal transmission of the first waveguide mode ($T_{11}$) between 6.65~GHz and 7.35~GHz. $T_{33}$ is shown in Fig.~\ref{fig:barrierSM}(b) for the third mode, and Fig.~\ref{fig:barrierSM}(c) displays the result for the last mode ($T_{44}$). 

A notable observation is the presence of strong spectral fluctuations in the transmission of the fourth mode. This behavior is attributed to the inherent properties of the mode itself—specifically, the larger time delay experienced in an empty waveguide due to the significant angle between the wavenumber $\mathbf{k}$ and the longitudinal direction \cite{horodynskiAntireflectionStructurePerfect2022}. As a result, the fourth mode exhibits heightened sensitivity to experimental uncertainties, such as slight variations in the positioning of antennas and scatterers.

\end{document}